\documentclass[twocolumn,showpacs,preprintnumbers,amsmath,amssymb]{revtex4-1}

\usepackage{graphicx}
\usepackage{dcolumn}
\usepackage{bm}
\usepackage{epsfig}

\usepackage{txfonts} 

\begin{document}

\title{Persistent electrical doping of Bi$_2$Sr$_2$CaCu$_2$O$_{8+x}$ mesa structures}

\author{Holger Motzkau}
\author{Thorsten Jacobs}
\author{Sven-Olof Katterwe}
\author{Andreas Rydh}
\author{Vladimir M. Krasnov}

\email{Vladimir.Krasnov@fysik.su.se}

\affiliation{Department of Physics, Stockholm University, AlbaNova
University Center, SE -- 106 91 Stockholm, Sweden }

\date{\today}

\begin{abstract}
Application of a significantly large bias voltage to small
Bi$_2$Sr$_2$CaCu$_2$O$_{8+x}$ mesa structures leads to persistent
doping of the mesas. Here we employ this effect for analysis of
the doping dependence of the electronic spectra of Bi-2212 single
crystals by means of intrinsic tunneling spectroscopy. We are able
to controllably and reversibly change the doping state of {\it the
same} single crystal from underdoped to overdoped state, without
changing its chemical composition. It is observed that such
physical doping is affecting superconductivity in Bi-2212 similar
to chemical doping by oxygen impurities: with overdoping the
critical temperature and the superconducting gap decrease, with
underdoping the $c$-axis critical current rapidly decreases due to
progressively more incoherent interlayer tunneling and the
pseudogap rapidly increases, indicative for the presence of the
critical doping point. We distinguish two main mechanisms of
persistent electric doping: (i) even in voltage contribution,
attributed to a charge transfer effect, and (ii) odd in voltage
contribution, attributed to reordering of oxygen impurities.
\end{abstract}


\maketitle

\section{Introduction}

High temperature superconductivity (HTSC) in cuprates occurs as a
result of doping of a parent antiferromagnetic Mott insulator and
properties of cuprates change significantly with doping.
Superconductivity in overdoped cuprates is fairly well described
by the conventional BCS-type second-order phase transition
\cite{TallonPhC,SecondOrder,MR}. But properties of underdoped
cuprates are abnormal due to the persistence of the normal state
pseudogap, strong superconducting fluctuations, or possibly
preformed pairing, \cite{Kaminski} and magnetism \cite{Kerr_Shen}
at $T>T_\text{c}$. There are indications that the transition from the
normal to the abnormal behavior occurs abruptly at a critical
doping point
\cite{TallonPhC,Doping,Balakirev,Kartsovnik,LeBoeuf2011}. This may
be a consequence of the quantum phase transition - a phase
transition, which occurs at $T=0$, in frustrated systems as a
result of a competition of coexisting order parameters. The coexistence
of superconductivity at $T<T_\text{c}$ with the pseudogap
\cite{KrasnovPRL2000,Lee2007,Bernhard}, charge and spin density
order \cite{Kartsovnik,LeBoeuf2011,Nernst2010} was indeed reported
by several techniques. Clearly, detailed doping-dependent studies are needed both for
understanding the puzzling nature of HTSC in cuprates and for the
development of novel HTSC materials.

Usually, the mobile carrier concentration is controlled by
chemical doping via chemical substitution, or in case of cuprates
also by variation of the oxygen content via appropriate annealing
and subsequent quenching to room temperature. This allows an
accurate control of the chemical composition, but less so of the
local arrangement of impurities and disorder, which is equally
important for cuprates \cite{DisorderBi2201}. For example, it is
well established that properties of the
$\mathrm{YBa_2Cu_3O_{6+x}}$ compound strongly depend not only on
the average oxygen concentration but also crucially on the
order/disorder of oxygen atoms in Cu-O-Cu chains
\cite{Chains,Raman}. Therefore, analysis of the doping phase
diagram of cuprates requires accurate control of both, the
concentration and the microscopic structure of impurities.

The carrier concentration can be also varied via two physical
doping processes, well established for semiconductors:
photo-doping \cite{Kudinov,Schuller,PhotoScience2011} and through
the electric-field effect
\cite{Ahn,Moeckly,Sorkin,Tulina,KovalCurrentInj,KovalMemres}. In
case of cuprates, physical doping may be persistent at low
temperatures in a sense that it is relaxing very slowly after
removing the light \cite{Kudinov,Schuller} or field
\cite{Moeckly,Tulina,KovalCurrentInj,KovalMemres}. Recently, a
persistent electric doping via strong current injection was
discovered \cite{KovalCurrentInj,KovalMemres}. It is resembling a
resistive switching phenomenon in memristor devices
\cite{Memristor1,Memristor2} and is related to previous similar
observations in point contact experiments on
Bi$_2$Sr$_2$CaCu$_2$O$_{8+x}$ (Bi-2212) \cite{Tulina}. Such an
electric doping is reversible, reproducible and easily
controllable. It opens a possibility to analyze the doping
dependence of HTSC on one sample without changing its chemical
composition \cite{PhotoScience2011}. Despite that, there were very
few direct spectroscopic studies of cuprates employing physical
doping techniques.

Intrinsic Tunneling Spectroscopy (ITS) provides an unique
opportunity to probe bulk electronic properties of HTSC \cite{SecondOrder}. This technique
utilizes weak interlayer ($c$-axis)
coupling in quasi two dimensional HTSC compounds, in which mobile
charge carriers are confined in CuO$_2$ planes separated by some
blocking layer (e.g. SrO-2BiO-SrO in case of Bi-2212). This leads
to a formation of atomic scale intrinsic tunnel junctions,
and to an appearance of the intrinsic Josephson effect at
$T<T_\mathrm{c}$
\cite{Kleiner94,KrasnovPRL2000,Suzuki,Katterwe2009,Superluminal}.

In this work we employ the persistent electric doping for analysis
of the doping dependence of electronic spectra of Bi-2212 single
crystals by means of intrinsic tunneling spectroscopy
\cite{SecondOrder,Doping,MR,KrasnovPRL2000,KatterwePRL2008}.
Controllable and reversible persistent physical doping is achieved
by applying a $c$-axis voltage of a few volts to small Bi-2212
mesa structures. Thus we are able to change the doping state of
Bi-2212 single crystals without changing its chemical composition.
A wide doping range from a moderately underdoped to strongly
overdoped state could be reached. It is observed that the physical
doping is affecting the intrinsic tunneling characteristics of
Bi-2212 similar to chemical doping \cite{Doping}. With overdoping
the critical temperature and the superconducting gap decrease.
With underdoping the pseudogap rapidly increases, indicative for
the presence of the quantum critical doping point in the phase
diagram, and the $c$-axis critical current density rapidly
decreases, indicating a progressively more incoherent interlayer
tunneling. We distinguish two main mechanisms of persistent
electric doping: (i) an even in voltage contribution, attributed
to a charge transfer effect, and (ii) an odd in voltage
contribution, attributed to reordering of oxygen impurities.

The paper is organized as follows. In Sec.~II we make a brief
overview of physical doping mechanisms of cuprates. Sec.~III
provides experimental details. In Sec.~IV we present the main
experimental results and in Sec.~V we discuss possible mechanisms
of persistent electric doping, followed by conclusions.

\section{Physical doping of cuprates}

\subsection {Photo-doping}

Photo-doping allows a wide-range variation of doping in the same
sample \cite{Kudinov,Schuller,PhotoScience2011}. The ordinary
non-equilibrium photo-doping is quickly relaxing because of a very
short life time ($\sim$ps) of photoinduced charge carriers
\cite{PhotoScience2011}. However, in underdoped $\mathrm{YBa_2Cu_3O_{6+x}}$ and some
other cuprates a different type of persistent photo-doping takes
place \cite{Kudinov,Schuller}. It involves significant energies
$\sim$$1\,\mathrm{eV}$, which makes it metastable at low
temperatures. Several mechanisms are contributing to the
persistent photo-doping \cite{PhDopTheory}, such as charge
transfer, which changes the redox state of the impurity atom
\cite{Kudinov}, and ordering of oxygen impurities in the lattice
\cite{Schuller}. Photo-doping always leads to an increase of the
doping level with respect to the initial state.

\subsection{Electric field effects}

Electric fields may both increase or decrease the number of
mobile charge carriers, depending on the direction of the applied
field \cite{Ahn}. The ordinary electric field effect is not
persistent and exists only during the time an electric field
is applied. Since the electric field penetrates only to the
Thomas-Fermi charge screening length, $\lambda_\mathrm{TF} \lesssim 1\,\mathrm{nm}$,
just a thin surface layer can be modified \cite{Shapiro}.

A persistent electrostatic field-effect due to net electric
polarization or trapped charges can be realized at the interface
between a superconductor and a ferroelectric \cite{SuperFerro} or
polar insulator \cite{ElStatic}. This is also a surface
phenomenon, but in case of layered cuprates, which represent
stacks of metallic CuO planes sandwiched between polar-insulating
layers \cite{Polariton}, electrostatic charging of insulating
layers may in principle lead to the {\it bulk} persistent
electrostatic field-effect.

Another type of a persistent and bulk electric-field effect has
been observed at large current densities \cite{Moeckly}.
Similar to photo-doping, it was attributed to a charge transfer
\cite{Salluzzo2008} and reordering of oxygen impurities
\cite{Sorkin}. Significant oxygen mobility in intense electric
fields also leads to a resistive switching phenomenon \cite{Tulina}.

\subsection{Resistive switching in complex oxides}

The resistive switching phenomenon occurs in many complex oxides
and is the basis for the development of resistive memory devices.
Several mechanisms may be involved in the resistive switching
phenomenon \cite{Memristor1,Memristor2}, including a change of the
redox state of some of the elements, oxygen migration, and
filament formation. Resistive switching has been observed on
depleted surfaces of Bi-2212 cuprates \cite{Tulina} and attributed
to oxygen migration. Recently, it was demonstrated that the
resistive switching-like phenomenon can be used for controllable
and reversible doping of small Bi-2212 micro-structures over a
wide doping range \cite{KovalCurrentInj,KovalMemres}.

\section{Experimental}

Mesas were fabricated on top of freshly cleaved Bi-2212 single
crystals by means of optical lithography, Ar ion milling and
focused ion beam trimming. Four batches of crystals were used:
pure near optimally doped (OP) Bi-2212, pure strongly underdoped (UD) Bi-2212, lead-substituted
$\mathrm{Bi_{1.75}Pb_{0.25}Sr_2CaCu_2O}_{8+\delta}$ [Bi(Pb)-2212],
and yttrium-doped $\mathrm{Bi_2Sr_2Ca_{1-x}Y_xCu_2O}_{8+\delta}$ [Bi(Y)-2212].
Mesas of different sizes from $5 \times 5$ to $1 \times
0.5~\muup\mathrm{m}^2$ and with a different number of junctions $N
= 8-56$ were studied. Details of the sample fabrication can be
found in Ref.~\cite{Submicron}. All studied mesas exhibited a
persistent doping effect upon application of a sufficiently large
bias voltage.

\begin{figure*}[t]
\begin{center}
\includegraphics[width=0.88\linewidth]{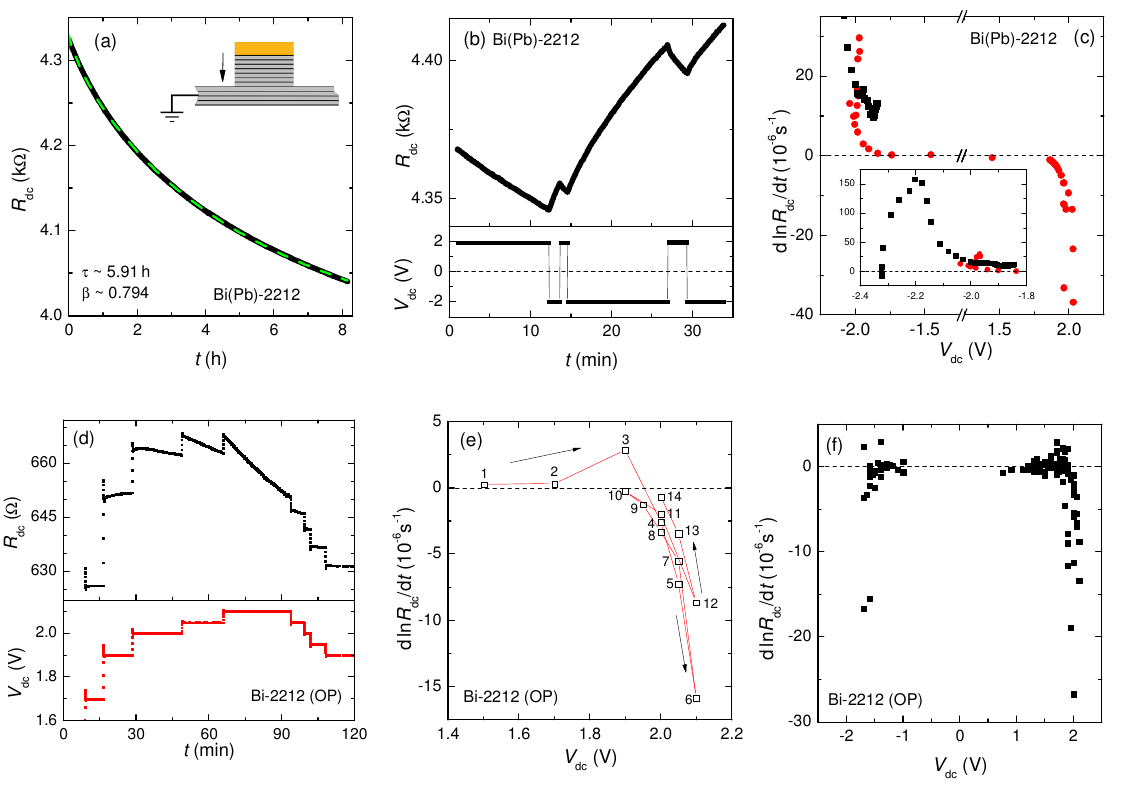}
\end{center}
\caption{(Color online). Dynamics of electric doping (a-c) for a
Bi(Pb)-2212 mesa at $T=135\,\mathrm{K}$ and (d-f) for an optimally
doped Bi-2212 mesa at $T=2\,\mathrm{K}$. Panel (a) shows time
evolution of the dc resistance at a bias of $\approx
1.9\,\mathrm{V}$. The dashed line represents a stretched
exponential decay. The inset schematically shows a sketch of a
mesa structure and the direction of electric field at positive
bias. (b) Demonstration of odd-in-voltage doping: the mesa
resistance increases/decreases at positive/negative bias. Panel
(c) shows the predominantly odd voltage dependence of the
logarithmic rate of the dc resistance change $\mathrm{d} \ln
R_\mathrm{dc}/\mathrm{d}t$. Different symbols represent different
runs. The inset in (c) demonstrates the sign change of the doping
direction at higher bias. Panel (d) demonstrates a gradual change
from a negative to positive doping rate with increasing bias
voltage and time. It also demonstrates a history dependence of the
doping rate, i.e., a different sign of resistance change at the
same bias voltage, depending on the former doping treatment. The
history dependence upon several sweeps of the bias voltage are
shown in (e). Panel (f) demonstrates the predominantly even in
voltage doping for OP Bi-2212. However an asymmetry indicates the
presence of a subdominant odd in voltage doping. Note that despite
a significant difference between the two mesas, the threshold
doping voltage is similar $\sim \pm 1.7\,\mathrm{V}$ (c). }
\label{fig:dynamics}
\end{figure*}

The samples were placed in a flowing gas cryostat and measured in
a three-probe configuration with a common top gold contact. The
ground contacts for current and voltage were provided through
other mesas on the same crystal. A Keithley K6221 current source
and a FPGA-based arbitrary waveform
generator and lock-in amplifier were used to bias and measure the
samples. Biasing was done at pseudo-constant voltage, with an
optional small superimposed ac voltage to simultaneously measure
the high-bias differential resistance in addition to the dc
resistance. Positive bias is defined as current going into the
mesa through the common top contact, as sketched in Fig.~\ref{fig:dynamics}\,(a).

\section{Results}

From chemical (oxygen) doping studies it is known that
doping/undoping of Bi-2212 is accompanied by a proportional
decrease/increase of the $c$-axis resistivity \cite{Doping}.
Therefore, we can control the doping
state by tracing the mesa resistance.

\subsection{Dynamics of electric doping}

The basic features of the dynamics of the persistent electric
doping are shown in Fig.~\ref{fig:dynamics}. Panel (a) shows the
time-evolution of the Bi(Pb)-2212 mesa resistance at a bias of
$\approx$$1.9\,\mathrm{V}$ at $T=135\,\mathrm{K}$. It is seen that
the mesa resistance is decreasing with time, indicating a gradual
doping of the mesa. The doping rate decreases
with time, following a stretched exponential
decay, $R = R_0+R_\mathrm{d}\exp[-(t/\tau)^\beta]$, shown by the
dashed line in in Fig.~\ref{fig:dynamics}\,(a), which is also
typical for persistent photo-doping \cite{Kudinov,Schuller}. The
doping can be equally well performed at any temperature
from 4 to 300\,K, but the rate is increasing with $T$. In most
cases we perform doping at low $T$ in order to be able to
immediately probe the superconducting characteristics. Upon
reduction of the bias below the threshold voltage the state of the
mesa remains stable even at room temperature on the time scale of
several days.

The resistive change is reversed upon voltage reversal, as
illustrated in Figs.~\ref{fig:dynamics}\,(b) and (c) for the same
Bi(Pb)-2212 mesa. The resistance decreases for positive bias
(electric field into the crystal) and increases for negative bias
(electric field towards the top contact), which indicates that we
can controllably and reversibly dope and undope the mesa. The
doping rate and direction depend both on the sign and the absolute
value of bias voltage.

\begin{figure*}[t]
\includegraphics[width=0.88\linewidth]{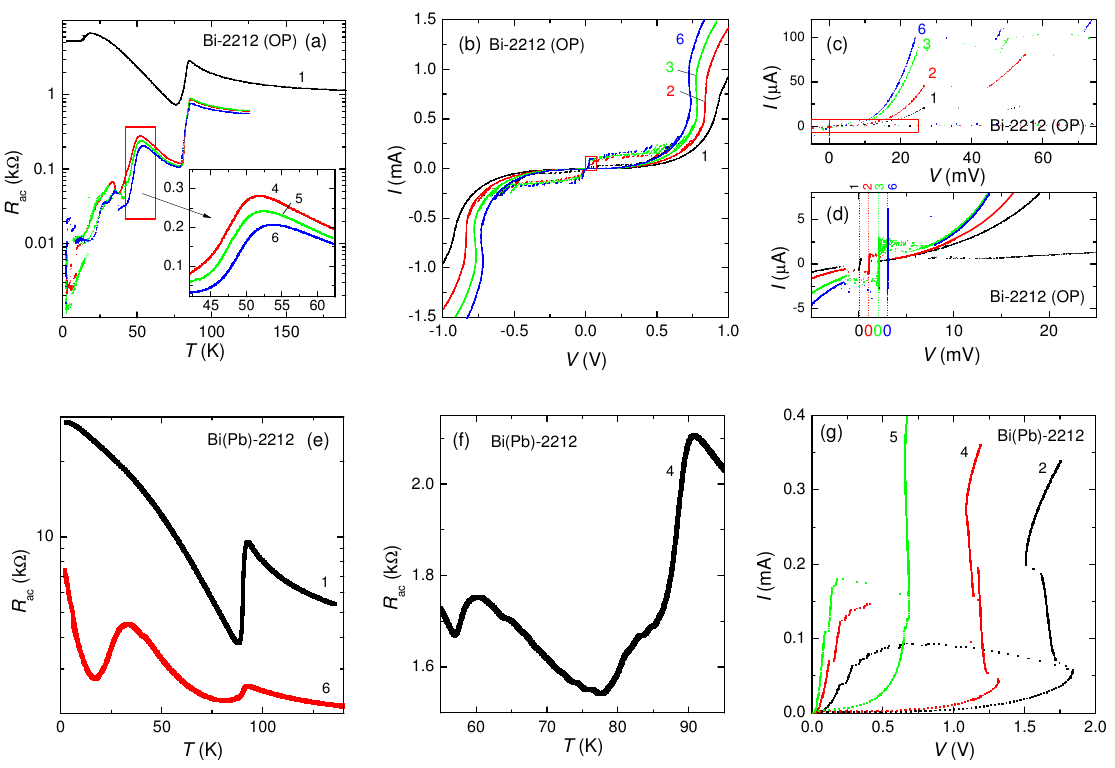}
\caption{(Color online) Intrinsic tunneling characteristics of the
same OP Bi-2212 (a-d) and Bi(Pb)-2212 (e-g) mesas as in Fig. 1 at
different doping states. (a) $T$ dependence of the ac resistance
$R_\text{ac}(T)$ in the initial high-resistive state (1) and
subsequent low resistance doping states (4-6). The inset
illustrates the significant enhancement of $T_\text{c}$ of the
depleted surface junction. Panels (b-d) represent different parts
of $I-V$ characteristics in the doping states 1, 2, 3 and 6 at
$T=2$ K. A variation of the tunnel resistance, sum-gap kink and
the critical current is clearly seen. Panels (e) and (f) show
$R_\text{ac}(T)$ for Bi(Pb)-2212 in the initial high-resistive
state (1) and subsequent low resistance doping states (4) and (6).
This mesa with a large number of junctions $N=56$ exhibits a
significant spread in the $T_\text{c}$ of individual junctions,
seen as small resistance drops in (f). Panel (g) shows $I-V$s of
the Bi(Pb)-2212 mesa in the doping states 2, 4 and 5. A
progressive increase of the critical current and decrease of the
sum-gap kink voltage is seen. } \label{fig:Tdep}
\end{figure*}

Figure~\ref{fig:dynamics}\,(c) summarizes the bias dependence of the doping rate
for the Bi(Pb)-2212 mesa. Below the threshold voltages, $\left|
V_\mathrm{dc}\right| \lesssim 1.7\,\mathrm{V}$, the mesa
resistance is stable. Upon increasing the bias voltage,
the resistance of the mesa starts to gradually change at a rate
that increases drastically up to $|V|\sim 2.2\,V$ as shown in the
inset of Fig.~\ref{fig:dynamics}\,(c). A further voltage increase
reduces the rate and then reverses the resistance alteration rate
(see the inset at $V=-2.32\,\mathrm{V}$). The sign change of the
alteration rate at high bias is in agreement with the observations
by Koval \emph{et\,al.} \cite{KovalCurrentInj}. The behavior in
this regime is, however, history dependent, as may be seen from
Fig.~\ref{fig:dynamics}\,(c), and the final state depends on how long time the
mesas was biased at every bias voltage. The doping process for the
Bi(Pb)-2212 mesa, Fig.~\ref{fig:dynamics}\,(c), is predominantly odd in bias
voltage, i.e., the direction of doping is changed when the sign of
the bias voltage is changed.

Figures~\ref{fig:dynamics}\,(d) and (e) show a detailed view of
the time and voltage dependence of doping for a near optimally
doped pure Bi-2212 mesa. The top panel of (d) shows the time
evolution of the resistance of an OP Bi-2212 mesa for different
bias voltages, shown in the bottom panel. It is seen that the
resistance is constant at $V=1.7\,\mathrm{V}$, and starts to
increase slowly at 1.9\,V. However, at 2\,V the resistance
initially increases but then starts to decrease after a few
minutes. This clearly shows that there are two counteracting
processes: a positive rate mechanism that saturates quickly and a
mechanism with negative rates that is dominating at longer times
and at higher voltages. The second process also saturates with
time, which is clear from Fig.~\ref{fig:dynamics}\,(a) and history dependent rates
of Fig.~\ref{fig:dynamics}\,(e). At larger voltage, the resistance
steadily decreases at a rate which is strongly bias dependent, as
shown in Fig.~\ref{fig:dynamics}\,(f).

Figures~\ref{fig:dynamics}\,(e,f) show the bias dependence of doping and a history
dependence upon sequential voltage sweeps (e). It is seen that for
the OP Bi-2212 mesa the electric doping is predominantly even in
voltage, i.e., the direction of doping does not depend on the sign
of the bias voltage. However, certain asymmetry of the doping rate
versus bias voltage characteristics in Figs. 2 (c) and (f)
indicates the presence of sub-dominant even- and odd-in-voltage
contributions for Bi(Pb)-2212 and OP Bi-2212 mesas, respectively.

\begin{figure*}
\begin{center}
\includegraphics[width=0.88\linewidth]{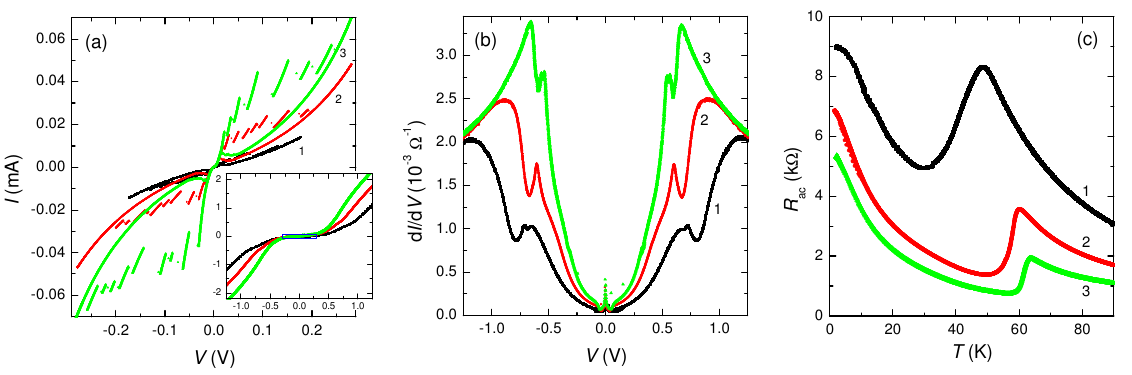}
\end{center}
\caption{(Color online). Doping of a strongly underdoped Bi-2212
mesa. (a) $I$-$V$ characteristics in the initial state 1 (black)
and two successive doping states 2 (red) and 3 (green). The
increase of the critical current by a factor five is seen. (b) The
corresponding $\mathrm{d}I$-$\mathrm{d}V$ characteristics. The
superconducting sum-gap peak moves to slightly lower voltages and
becomes sharper with doping. The $c$-axis pseudogap hump voltage
rapidly decreases with doping. (c) Zero bias ac resistivity
$R_\text{ac}(T)$. The $T_\text{c}$ has increased by about 15\,K in the state
3.} \label{Thorsten}
\end{figure*}

\subsection{Doping dependence of ITS characteristics }

Using the described method, the superconducting properties of the
mesas have been altered to different intermediate doping states
denoted by a successive number. The electric doping changes all
mesa characteristics: the $c$-axis resistivity, the critical
temperature $T_\mathrm{c}$, the $c$-axis critical current density
$J_\mathrm{c}$, the superconducting energy gap, $\Delta$, the
$c$-axis pseudogap and the $c$-axis resistivity in a manner very
similar to chemical (oxygen) doping \cite{Doping}.

Figure~\ref{fig:Tdep}\ shows temperature dependencies of low bias
ac-resistances for the initial, high resistance state (HRS), and
doped, lower resistance states (LRS), for (a) an OP Bi-2212 and
(e,f) a Bi(Pb)-2212 mesa. The $I$-$V$ characteristics of those
mesas at different doping states are presented in panels (b-d) and
(g), respectively.

From Fig.~\ref{fig:Tdep}\,(a) it is seen that the initial state
was characterized by the strong thermal-activation-type increase
of resistance with decreasing $T$, typical for underdoped Bi-2212
\cite{Katterwe2009}. The general shape of the resistive transition
was described in Ref.~\cite{SecondOrder}. At $T_\mathrm{c} \sim
82\,\mathrm{K}$ the resistance dropped to the top-contact
resistance, which originates from the first deteriorated junction
between the top CuO plane, shortly exposed to atmosphere after
cleavage, and the second, un-deteriorated CuO plane
\cite{SecondOrder}. Initially this junction had a very low
$T_\mathrm{c}'$ and a very small critical current, $I_\mathrm{c}$,
as can be seen from the corresponding $I$-$V$ in panel (d). After
electric doping, the resistance in the normal state dropped almost
three times and became less semiconducting. The main
$T_\mathrm{c}$ of the mesa changed only slightly, indicating that
the doping was changing around the optimal doping level with the
flat $T_\mathrm{c}$ vs. doping dependence. However, the properties
of the top junction changed drastically: the $T_\mathrm{c}'$
increased to $\sim 50\,\mathrm{K}$, and the critical current
increased ten-fold as shown in panel (d), even though it still
remains $\sim 20$ times smaller than for the rest of the junctions
in the mesa, as can be seen from panel (c). This indicates that
the surface CuO plane was initially strongly underdoped and,
therefore, responded much stronger to variation of doping, due to
the steep $T_\mathrm{c}$ vs. doping dependence at the underdoped
side of the doping phase diagram of cuprates \cite{Doping}. A
similar trend was also observed in photo-doping
\cite{Kudinov,Schuller}.

\begin{figure*}
\begin{center}
\includegraphics[width=0.88\linewidth]{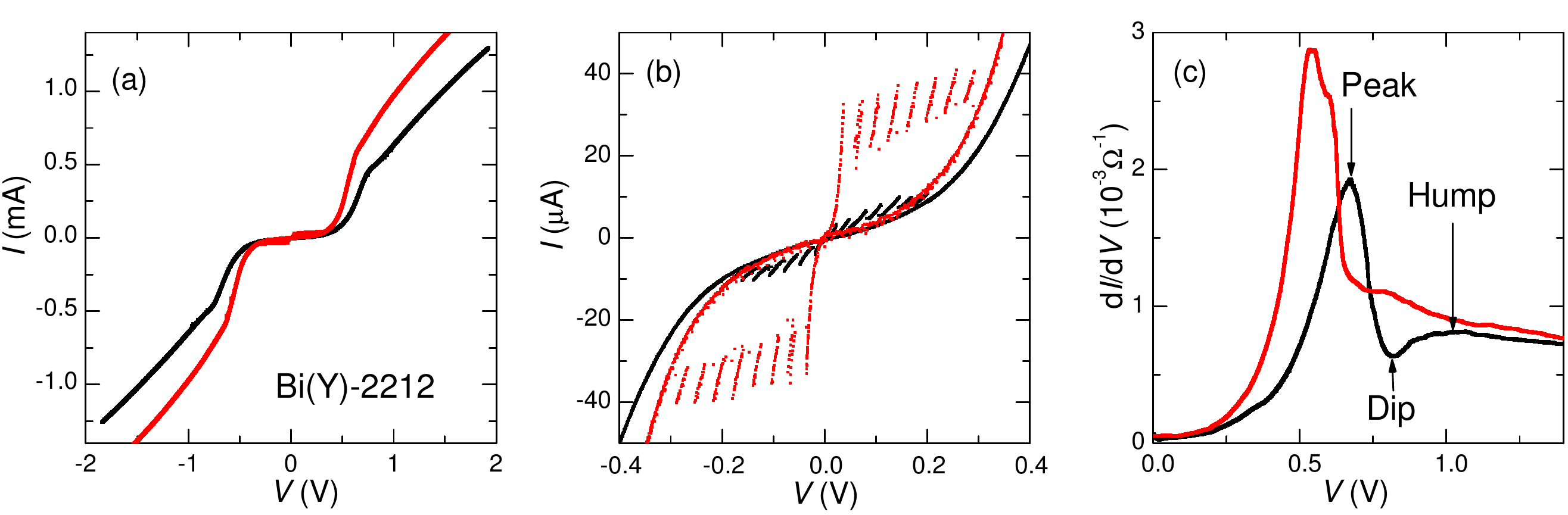}
\end{center}
\caption{(Color online). Intrinsic tunneling characteristics of a
small Bi(Y)-2212 mesas at low $T$ before (slightly underdoped,
HRS) and after (optimally doped, LRS) doping by a short voltage
pulse. Panels (a) and (b) show $I$-$V$ curves at a large scale and
at the quasiparticle branches, respectively. A reduction of the
high bias tunnel resistance and a simultaneous increase of the
critical current in the doped state is seen. Panel (c) shows
$\mathrm{d}I/\mathrm{d}V$ ITS characteristics. A change in the
shape of the curves is seen: the peak becomes sharper and the
dip-hump less pronounced in the optimally doped state, compared to
the initial underdoped state. } \label{IVSMa1b}
\end{figure*}

Another possible reason for the stronger response of the
underdoped top junction is the larger $c$-axis resistivity of
underdoped intrinsic Josephson junctions \cite{Doping}.
Because of that the electric field is not uniformly distributed
along the mesa but is larger in the high-resistive top junction.
This together with the strong voltage dependence of the electric
doping leads to a faster doping of the top junction and the doping
may even go in the
opposite direction with respect to the rest of the mesa.

The effect of non-uniform doping along the height of the mesa
becomes more pronounced in higher mesas with a larger number of
intrinsic Josephson junctions. This is seen from $R(T)$ for the
Bi(Pb)-2212 mesa from Fig.~\ref{fig:Tdep}\,(e), which contained a
fairly large number of junctions $N\approx 56$. It is seen that
after doping some junctions retained the initial $T_\mathrm{c}
\sim 90\,\mathrm{K}$, but some were very strongly overdoped to
$T_\mathrm{c} \sim 30\,\mathrm{K}$. Fig.~\ref{fig:Tdep}\,(f) shows
$R(T)$ at the intermediate doping state 4. Small drops represent
critical temperatures of individual junctions in the mesa.
Apparently there is a gradual distribution of $T_\mathrm{c}$ along
the height of the mesa.

Doping of Bi-2212 leads to a rapid increase of the $c$-axis
critical current density \cite{Doping}. This is clearly seen from
$I$-$V$ curves for the Bi-2212 (OP) mesa shown in
Figs.~\ref{fig:Tdep}\,(b-d). The increase of $I_\mathrm{c}$ of the
surface junction is shown in panel (d). One-by-one switching of
the rest of the junctions into the resistive state at
$I>I_\mathrm{c}$ leads to the appearance of multiple-quasiparticle
(QP) branch structures in the $I$-$V$ curves. The corresponding
critical current at the first QP branch is shown in panel (c). It
also strongly increases with doping. The same effect doubles
$I_\mathrm{c}$ in the Bi(Pb)-2212 mesa, in Fig.~\ref{fig:Tdep}
(g).

From Fig.~\ref{fig:Tdep}\,(b) it is seen that the $I$-$V$ curves
exhibit a kink at large bias, followed by an ohmic tunnel
resistance. The kink represents the sum-gap singularity in
superconducting tunnel junctions at $V=2\Delta/e$ per junction
\cite{SecondOrder,MR}. This is the basis of the ITS technique,
which allows analysis of the superconducting energy gap $\Delta$
in the bulk of the Bi-2212 single crystal.

Accurate analysis of the electronic spectra with the ITS technique
requires mesas with a small area and a small number of identical
junctions. This is needed for avoiding possible artifacts,
associated with self-heating, in-plane non-equipotentiality, and
spread in junction parameters \cite{Heating_PRL2005,SecondOrder}.
This is particularly important for the analysis of the genuineshape of tunneling characteristics, which remains a controversial
issue \cite{Comment2}. Even though the $I$-$V$ characteristics in
Figs.~\ref{fig:Tdep}\,(b) and (g) are distorted by self-heating, as evident
from a back-bending at large bias, the general trend for a
variation of the sum-gap kink with doping is clearly seen: The
superconducting gap decreases and the sum-gap kink becomes sharper
with \mbox{(over-)doping}. This qualitative conclusion is not
affected by self-heating because the dissipation power at the kink
decreases with decreasing resistance and becomes smaller with
subsequent doping. Thus, with over-doping the superconducting
sum-gap singularity becomes sharper and moves to lower voltages
despite the progressive reduction of self-heating. This clearly
reveals the doping-variation of the genuine $c$-axis tunneling
characteristics \cite{Comment2}. A similar tendency was observed
by other techniques, including the angular resolved photoemission
spectroscopy \cite{ARPES_doping}, scanning tunneling spectroscopy
\cite{STS_inhomo}, and tunneling spectroscopy on point contacts
\cite{Zasad_doping}, as well as in previous ITS studies involving
chemical (oxygen) doping
\cite{Doping,KatterwePRL2008,SecondOrder}.

Figure~\ref{Thorsten} shows the electric doping of an initially strongly
underdoped Bi-2212 mesa. A five-fold increase in critical current
in the $I$-$V$ characteristics (a), and an increase of
$T_\mathrm{c}$ of about 15\,K (c) can be seen. The tunneling
conductance $\mathrm{d}I/\mathrm{d}V(V)$ curves are shown in
panel (b). It is seen that the sum-gap peak shifts to slightly
lower voltages and becomes sharper with doping. The hump voltage,
attributed to the $c$-axis pseudogap \cite{SecondOrder}, rapidly
decreases with increasing conductance, pointing towards an abrupt
opening of the pseudogap at the critical doping point, consistent
with chemical doping studies \cite{Doping,Balakirev,Bernhard}.

Note that in all studied cases the electric doping has lead to a
significant increase of the critical current, while the
superconducting gap was decreasing. The anticorrelation between
$I_\text{c} R_\text{n}$ and $\Delta$ in underdoped Bi-2212 has been reported
before \cite{Doping} and was attributed to progressively more
incoherent $c$-axis transport in combination with the $d$-wave
symmetry of the superconducting order parameter.

\subsection{Short-pulse doping}

\begin{figure*}[t]
\includegraphics[width=0.88\linewidth]{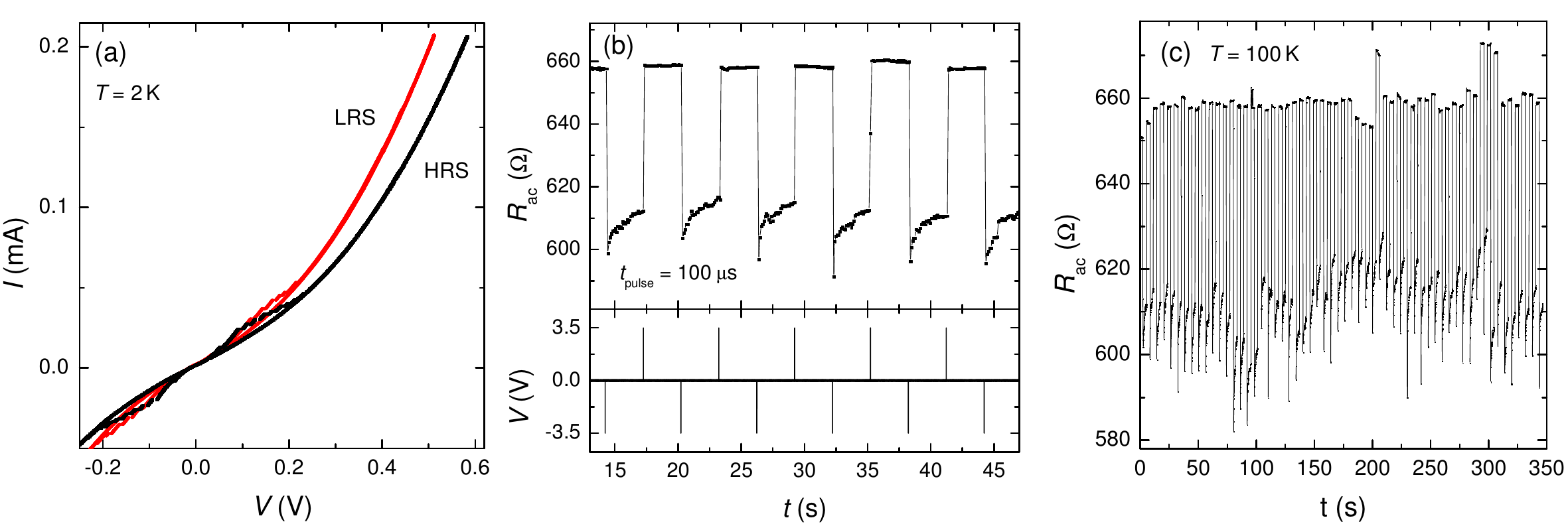}
\caption{(Color online) Short pulse resistive switching in a
strongly underdoped Bi-2212 mesa. (a) $I$-$V$ characteristics in
the initial high-resistive state (HRS) and the doped low-resistive state (LRS) at $T=2$\,K.
Panels (b) and (c) demonstrate the resistive switching sequence
$R_\text{ac}$ versus time at $T=100$\,K. The switching was made by
a train of short positive and negative pulses, shown in the bottom
of panel (b). It is seen that the HRS is stable, but the LRS is
initially relaxing and then saturates at a resistance below the
HRS. Several hundred reproducible resistive switching events can
be achieved without visible degradation (c).} \label{fig:pulse}
\end{figure*}

So far we were discussing a gradual electric doping of Bi-2212
mesas at the time scale of an hour, as shown in Fig.~\ref{fig:dynamics}.
Such a long time doping allows a very strong variation of the doping state, but
often leads to inhomogeneous doping within the mesa height, as shown in Fig.~\ref{fig:Tdep}.
Koval et.al.~\cite{KovalMemres} demonstrated that a short-pulse doping
strategy leads to highly reversible and reproducible doping, similar to resistive
switching in point contacts \cite{Tulina}.
This is probably related to the lack of significant
electromigration during the short pulse, which may eventually lead
to an irreversible destruction of the crystal structure
\cite{Moeckly}.

Figure~\ref{IVSMa1b} represents the ITS characteristics at $T \sim
30\,\mathrm{K}$ for a small $\sim 2\times 2\,\mathrm{\muup m^2}$
Bi(Y)-2212 mesa with a small amount of junctions and small
self-heating \cite{Heating_PRL2005}. In the initial high-resistive
state the mesa is slightly underdoped with $T_\mathrm{c}\sim
91\,\mathrm{K}$. The mesa was switched to a low-resistive state by
a short voltage pulse $V \gtrsim 2\,\mathrm{V}$ of about a
milliseconds width. The periodicity of QP branches in
Fig.~\ref{IVSMa1b}\,(b) demonstrates that after switching into the
LRS the mesa remains highly uniform. From comparison of $I$-$V$
curves in panels (a) and (b) it is seen that the decrease of
resistance by $\sim 1/3$ is accompanied by an almost four-fold
increase of $I_\mathrm{c}$. The $T_\mathrm{c}$ increased to $\sim
93\,\mathrm{K}$ being an indication that the mesa became near
optimally doped.

Fig.~\ref{IVSMa1b}\,(c) represents the tunneling conductance
$\mathrm{d}I/\mathrm{d}V$ in the high-resistive (underdoped) and
the low-resistive (optimally doped) states. The following main
changes in ITS spectra are seen: the superconducting sum-gap peak
voltage decreased in the doped low resistive state
and the shapes of spectra are changed. The relative sum-gap peak
height $\mathrm{d}I/\mathrm{d}V(V_\mathrm{p}) R_\mathrm{n}$ is
increased by about 50\% in the low-resistive
state.
The high-resistive state exhibits a peak-dip-hump structure, which
is less obvious in the low-resistive state. All this is similar to
the slow-doping case, Figs.~\ref{fig:Tdep} and 3, and consistent
with the change of doping from the slightly underdoped to near
optimally-doped state \cite{Doping}, in accordance with other
spectroscopic studies involving chemical (oxygen) doping
\cite{ARPES_doping,STS_inhomo,Zasad_doping}.

As already discussed above, the increase of sharpness of the
sum-gap kink in the $I$-$V$ curve and the peak in the $\mathrm{d}I/\mathrm{d}V$ curve in
the low-resistive state, reported in Figs.~\ref{fig:Tdep} and
\ref{IVSMa1b}, and the change of the shape of the peak-dip-hump
feature in Fig.~\ref{IVSMa1b} (c) can not be attributed to
self-heating, because the actual power dissipation at the peak is
decreasing in the low-resistive state. For example, the
corresponding powers at the peaks in Fig.~\ref{IVSMa1b}\,(c) are
$P=0.21$ and 0.17\,mW, respectively. This observation supports the
conclusion of Ref.~\cite{Comment2} that the appearance and the
shape of the peak-dip-hump structure in the ITS characteristics of
small mesas is determined primarily by the doping level.

A detailed analysis of short pulsed resistive switching has been
performed on strongly underdoped Bi-2212 mesas. For negative pulses with
$50\,\mathrm{\muup{}s}$ length and compliance voltages up to
$-2.5\,\mathrm{V}$ there was no resistive switching. At
$-3.0\,\mathrm{V}$, a reduction of the quasiparticle resistance to
99.3\% is observed (LRS). Nine subsequent
pulses reduce the resistance further to 98.4\%. A single
positive pulse switches the resistance back
to the initial HRS, while more positive pulses do not result in additional changes. A single
negative pulse with a higher compliance voltage of
$-3.5\,\mathrm{V}$ instead reduces the resistance of the LRS to
about 93.4\% of the HRS, and the corresponding positive
pulses switches it back. Pulses with lower compliance voltage lead
to a partial switching
but additional pulses with the same compliance do not lead to a
significant change. Fig.~\ref{fig:pulse}\,(a) shows $I$-$V$ curves of another UD Bi-2212 mesa at $T=2\,\mathrm{K}$ in the HRS
and the LRS obtained with a $3.5\,\mathrm{V}$ pulse of $50\,\mathrm{\muup{}s}$ width. The general difference
between $I$-$V$s after the pulsed doping in Figs.~\ref{IVSMa1b}\,(a) and ~\ref{fig:pulse}\,(a) is the same as for the slow doping in Figs.~\ref{fig:Tdep}\,(a) and 3\,(a).

Fig.~\ref{fig:pulse}\,(b) demonstrates a reproducible resistive switching
between HRS and LRS at elevated $T=100\,\mathrm{K}$ for another UD
Bi-2212 mesa. The switching was performed using a similar positive
and negative pulse sequence with $\pm 3.5\,\mathrm{V}$, a pulse
width of $100\,\mathrm{\muup{}s}$ and an interval between pulses
of 3\,s, shown in the bottom panel of Fig.~\ref{fig:pulse}\,(b). Panels (b) and
(c) show the corresponding time sequence of the measured zero-bias
resistance of the mesa. It is seen that negative voltage pulses
lead to switching into the LRS while a subsequent positive pulse
switches the mesa back into the HRS. The corresponding resistance
change rates are of the order of $\mathrm{d}\ln
R/\mathrm{d}t\approx \mp 1000\,\mathrm{s^{-1}}$, which are
somewhat higher than for the slow doping shown in Fig.~\ref{fig:dynamics}\,(d), but not
inconsistent with that data, taking into account that the
compliance voltage is also significantly higher. It is seen that
the HRS is stable and shows no visible relaxation at
$T=100\,\mathrm{K}$, while the LRS is initially relaxing with the
characteristic time $\tau_\mathrm{LRS}=(0.77\pm 0.30)\,\mathrm{s}$
and then saturates before reaching the HRS, as shown in panel (b).
At $T=280\,\mathrm{K}$, the behavior is similar with a time
constant of $\tau_\mathrm{LRS}=(0.75\pm 0.34)\,\mathrm{s}$.

\section{Discussion: mechanisms of persistent electric doping}

In Sec.~II we briefly reviewed known mechanisms of persistent
physical doping of cuprates. It is likely that some of them are
playing a role in persistent electric doping, studied here.
Indeed, the phenomenon is clearly related to the persistent
electric-field effect \cite{Moeckly}, observed in
$\mathrm{YBa_2Cu_3O_{6+x}}$, which in turn is clearly related to
persistent photo-doping observed for various cuprates
\cite{Kudinov,Schuller}.

To identify possible mechanisms, we first summarize characteristic
features of the persistent electric doping:

i) Observed different voltage dependencies (odd and even) indicate that several distinct
mechanisms are involved.

ii) The doping rate shows a threshold-like behavior as a function
of bias voltage (see Fig.~\ref{fig:dynamics}). The threshold voltage depends on $T$
in a thermal-activation manner, i.e. decreases with increasing
$T$. Remarkably, the threshold voltage is weakly dependent on the
number of junctions in the mesa, consistent with previous reports
\cite{KovalCurrentInj,KovalMemres}, and is comparable to that for
a single point contact \cite{Tulina}. This suggests that the
phenomenon is connected to some characteristic energy $\sim
1\,\mathrm{eV}$, rather than directly to the electric field. Indeed,
for a given voltage, the latter should scale inversely
proportional to the number of junctions in the mesa, i.e., would
not be universal for different mesas (that is why we hesitate to
refer to the phenomenon as a persistent electric \textit{field} effect,
and rather call it persistent electric doping).

iii) However, the role of the electric field should not be
underestimated. Indeed, the displacement field is given by
$D=V\varepsilon_\mathrm{r}/Nt$, where $\varepsilon_\mathrm{r}$ is
the dielectric constant and $t$ is the thickness of the insulating
barrier between CuO bi-layers. The ratio $t/\varepsilon_\mathrm{r}
\simeq 0.1\,\mathrm{nm}$ was estimated from an analysis of Fiske
(geometrical resonance) step voltages \cite{Superluminal}.
Therefore, the displacement field in intrinsic junctions is

\begin{equation}\label{Field}
D \simeq \frac{V}{N}\times 10^8\,\mathrm{cm^{-1}}.
\end{equation}

For $V \simeq 2\,\mathrm{V}$ and $N\simeq 10$, which corresponds to the case
of Fig.~\ref{IVSMa1b}, one would get a very large value $D\simeq
2\times 10^7\,\mathrm{V/cm}$, which is certainly capable of seriously
polarizing and displacing ions in complex oxides
\cite{Moeckly,Tulina,Memristor1}.

iv) The phenomenon is not associated with a net change of the oxygen
content, which may only decrease within the cryostat. To the
contrary, mesas can be repeatedly and reversibly doped and
undoped, as shown in Figs.~\ref{fig:dynamics}\,(b) and ~\ref{fig:pulse}\,(b).

\subsection{Charge transfer and electrostatic charge trapping}

If an injected electron has a high enough energy, it may join one
of the ions, leading to a change of the redox state and the
effective doping. It was suggested that such a ``charge
transfer" mechanism is involved both in photo-doping of cuprates
\cite{Kudinov} and in the resistive switching phenomenon in other
complex oxides \cite{Memristor1}.

Alternatively, the electron may be trapped (localized) in
dielectric parts, leading to electrostatic charging of the sample.
The Bi-2212 compound has a layered structure with metallic CuO
planes sandwiched between polar insulator BiO layers
\cite{Polariton}. In this case the electrostatic charging will
take place in BiO layers, which may affect the doping state of the
neighboring CuO planes via the electrostatic field-effect. The
electrostatic charging of means by current injection takes place
uniformly within the whole structure. Therefore, unlike the
conventional electric field effect \cite{Ahn} and the
electrostatic field effect at the interface between a
superconductor and a ferroelectric material \cite{SuperFerro,ElStatic},
the current injection may lead to a persistent {\em bulk}
electrostatic field-effect doping of Bi-2212. Koval et.\,al.
\cite{KovalCurrentInj} emphasized the similarity of the phenomenon
with the floating-gate effect utilized in Flash memory devices.

Both types of charging effects have common similarities:

i) The charge transfer requires a certain energy ($\sim$eV for
$\mathrm{YBa_2Cu_3O_{6+x}}$), rather than electric field.

ii) The sign of the current and the direction of the electric
field does not matter. Therefore, such doping should be even with
respect to the voltage sign, consistent with the observations in Ref.~\cite{KovalCurrentInj}.

Therefore, we attribute even in voltage persistent electric doping to charge transfer and/or
charge trapping mechanisms.

\subsection{Oxygen reorientation and reordering}

It is well established that the doping state of cuprates depends
not only on the amount of off-stoichiometric oxygen, but also on
the relative orientation of the oxygen bonds \cite{Chains}. Therefore,
the doping state can be changed by oxygen reordering. Since the
required energy is large $\sim$$\mathrm{eV}$, compared to thermal
energies, oxygen reordering is a slow process and does not take
place spontaneously at low enough $T$. Oxygen reordering is
considered as one of the main mechanisms of the persistent photo
\cite{Schuller} and electric field \cite{Moeckly} doping.

In the case of persistent electric-field doping, the oxygen
reordering is steered by the polarization. Therefore, the
direction of doping should depend on the direction of the electric
field, i.e., should be odd with respect to the bias voltage. We
clearly see such a contribution in our experiment, see
Fig.~\ref{fig:dynamics}\,(b). Note that the odd-in-voltage contribution was
reported in the point-contact case \cite{Tulina}, but not reported
in previous related works made on zig-zag type Bi-2212
microstructures \cite{KovalCurrentInj,KovalMemres}.
The geometry of the latter samples is symmetric with respect to
the electric field direction (changing the sign of the electric
field is equivalent to flipping their sample up-side down). This
is not the case in point contacts and mesa structures, studied
here, for which the fields down (into the crystal) and up (into
the top electrode) are not equivalent. Therefore, the difference
may partly be due to the difference in sample geometry, or to the
observed sample-dependence of the relative strength of odd and
even in voltage doping contributions, as shown in Figs. 1 (d) and
(e).

Thus, we attribute the odd in voltage persistent electric doping
mechanism to field-induced oxygen reorientation/reordering.

\subsection{Irreversible processes: electromigration, filament and arc formation}

An increase of the bias voltage above $\sim 2.5-3.0\,\mathrm{V}$
leads to a gradual increase of the current and irreversible change
of the mesa properties. A similar phenomenon was observed in
$\mathrm{YBa_2Cu_3O_{6+x}}$ thin films and attributed to
electromigration and field-induced diffusion of oxygen, which is
even in bias voltage. The increase of conductance is probably due
to a dielectric breakthrough in the insulating BiO layers, which
leads to a pin-hole and filament formation. Thus, we attribute the
slow and irreversible drift of the mesa characteristics at large
bias voltages to electromigration in the mesas. This destructive
process is, however, distinctly different from the reversible and
reproducible electric doping effect, reported above.

After deterioration by electromigration, the mesa characteristics
become similar to resistive switching characteristics for point
contacts on top of oxygen-depleted, Bi-2212 surfaces
\cite{Tulina}.
At even higher bias the resistance becomes very high (infinite).
But an inspection in a microscope shows that the mesa itself
remains intact. There is no physical evaporation of material or a
crater at the place of the mesa, as in the case of a violent
electric discharge. Instead there are clear indications of an arc
formation at one of the sharp corners of the mesa, which probably
leads to delamination of the structure and mechanical
disattachment of the mesa from the base crystal.

\subsection{Mechanisms of energy accumulation}

The most puzzling property of the reported persistent electric
doping is that the required bias voltage is weakly dependent on
the number of junctions in the mesa
\cite{KovalCurrentInj,KovalMemres}. This is clearly seen from the
presented data, for which the threshold voltage is always $\sim
2\,\mathrm{V}$ for $N=9$ in Fig.~\ref{IVSMa1b}, $N=11$ in Fig.~\ref{fig:dynamics}\,(f) and
$N\sim56$ in Fig.~\ref{fig:dynamics}\,(d), which is also similar to that for a
single point-contact $\sim 1.5$~V \cite{Tulina}. It is, therefore, clear that
electric doping requires a certain electron energy, rather than
electric field. However, for stacked tunnel junctions, the energy
acquired by the injected electron in every tunneling event is
proportional to the voltage drop across the junction, $\delta E
\simeq eV/N$, and is significantly smaller than the threshold
energy. The main question is, therefore, how the electrons
accumulate a sufficiently large energy, required for doping.

In Ref.~\cite{KovalMemres} it was suggested that an electron can
accumulate energy upon sequential tunneling through several
junctions without relaxation. However, in this case only electrons
in the last junction will have enough energy. This would result in
a strongly inhomogeneous doping in different junctions. Moreover, the
probability of sequential tunneling without relaxation is small,
because of the small ratio of relaxation time ($\tau
\sim\mathrm{ps}$) \cite{Cascade} to the tunnel time,
$t_\mathrm{tun}$, $t_\mathrm{tun} /\tau \gg 1$. The probability of
sequential tunneling through $N$ junctions is decreasing rapidly
$\propto (\tau /t_\mathrm{tun})^{N}$ with increasing $N$. This
should lead to a dramatic increase of the doping time with
increasing $N$. Indeed, suppose that it takes
$t_N \sim 10\,\mathrm{min}$ for $N$ junctions at $I=1\,\mathrm{mA}$ to dope the
mesa. This will involve $N_e = I t_N / e$ tunneling events in each
junction ($e$ is the electron charge). Since the probability of
sequential tunneling through $2N$ junctions is decreasing
quadratically, it would require $N_e^2$ tunneling events per
junction, which will take $t_{2N} = N_e t_N = 3.75 \times 10^{19}\,\mathrm{min}$.
However, such a dramatic increase of the doping time with increasing mesa
height is inconsistent with experiments.

For the sequential tunneling scenario to be relevant, the ratio
$t_\mathrm{tun} /\tau$ should rapidly drop with increasing
electron energy and become of the order of unity at $E \sim
1\,\mathrm{eV}$. This may be caused by resonant tunneling, which
increases the tunneling rate of quasiparticles with a certain
energy, and/or by a drastic slowing down of the high energy
quasiparticle relaxation, which may be caused by a rapid decrease
of the Eliashberg's electron-boson spectral function and a gap in
the corresponding bosonic density of states at high energies
\cite{Cascade}. Such a scenario is interesting to investigate
because it may give an information about the bosonic spectrum,
involved in Cooper pairing, and thus provide a clue about the
electron-boson coupling mechanism, responsible for
high-$T_\mathrm{c}$ superconductivity in cuprates \cite{Cascade}.

We also want to propose an alternative mechanism for the energy
accumulation of electrons:
the formation of electric-field domains in the natural
atomic superlattice formed by the mesa. Electric field domains are
well studied in semiconducting superlattices \cite{Superlattice}.
They appear in weakly coupled superlattices close to the resonant
tunneling condition. The corresponding non-linearity leads to an
instability and a multiple-valued current-voltage characteristics.
As a result, the electric field distribution in the superlattice
becomes nonuniform and is concentrated in one or several
junctions.

A possibility of a formation of electric field domains in Bi-2212
mesas is not just a hypothesis. In fact, the multiple-branch
$I$-$V$ of Bi-2212 mesas due to one-by-one switching of intrinsic
junctions from the superconducting to the resistive state, shown
in Fig.~\ref{IVSMa1b}\,(b), is due to a formation of electric field domains in
individual tunnel junctions. The formation of electric field domains
in Bi-2212 mesas at high bias would explain many of the features
of the studied persistent electric doping. In this case electrons
in the domain may gain an energy close to eV without sequential
tunneling through the whole mesa. Furthermore, since domains are
typically dynamic and propagate through the whole superlattice
\cite{Superlattice}, this would also explain the uniformity of doping
in the whole mesa, and not just in the outermost junction.

\section{Conclusions}

We have studied the effect of persistent electric doping on
intrinsic tunneling characteristics of small Bi-2212 mesa
structures. It was shown that the application of a sufficiently
large voltage to the mesas leads to a controllable and reversible
physical doping of the mesas, without a modification of their
chemical composition. This allows the analysis of bulk electronic
spectra in Bi-2212 in a wide doping range on one and the same
mesa. This physical doping has the same effect as chemical
(oxygen) doping on the intrinsic tunneling characteristics of
Bi-2212: the $c$-axis resistivity decreases, the critical current
increases and the energy gap is decreasing together with
$T_\text{c}$ with over-doping. The anticorrelation between
$I_\text{c} R_\text{n}$ and $\Delta$ indicates that the $c$-axis
transport becomes progressively more incoherent at the underdoped
side of the phase diagram. An analysis of the doping variation of
the intrinsic tunneling characteristics of the same mesa provides
a clue about its genuine shape: with subsequent doping, the
sum-gap peak in the tunneling conductance becomes sharper and the
pseudogap hump rapidly decreases with doping, suggesting the
presence of a critical doping point, in agreement with previous
chemical doping studies \cite{Doping}.

By analyzing the bias- and time dependence we could identify different
mechanisms involved in the persistent electric doping: i) The even-in-voltage process
via charge transfer and/or charge trapping. ii)
The odd-in-voltage process via oxygen reordering. Those are distinct from the
irreversible electromigration and oxygen electrodiffusion,
observed at higher bias.

We confirm the previous report \cite{KovalMemres} that the
threshold voltage for the electric doping is weakly dependent of
the number of junctions in the mesas and is similar to that for a
single surface point contact \cite{Tulina}. This indicates that it
is the energy of injected electrons, rather than electric field,
that determines the phenomenon. We suggest that the required
energy accumulation by tunnel electrons may be due to a formation of
electric field domains in the natural atomic superlattice formed in
the Bi-2212 single crystal.

{\bf Acknowledgments} We are grateful to the Swedish Research
Council and the SU-Core Facility in Nanotechnology for financial
and technical support.



\end{document}